# Розробка фільтру SageMath для Moodle

Євгеній Олександрович Модло[*], Сергій Олексійович Семеріков[≠]
Криворізький металургійний інститут
ДВНЗ «Криворізький національний університет»,
вул. Революційна, 5, м. Кривий Ріг, 50006, Україна
modea@mail.ru[*], semerikov@gmail.com[≠]

**Анотація.** *Цілі дослідження*: визначення особливостей процесу розробки, встановлення, налаштування та використання фільтру SageMath для системи підтримки навчання Moodle.

*Завдання дослідження*: обґрунтувати доцільність використання системи Moodle як засобу підтримки процесу формування у майбутніх бакалаврів електромеханіки компетентностей з моделювання технічних об'єктів; проаналізувати існуючі засоби підтримки діяльності з моделювання технічних об'єктів та визначити шляхи їх інтеграції з системою Moodle; описати структуру та особливості програмної реалізації нового фільтру SageMath для системи Moodle; надати рекомендації з встановлення та налаштування розробленого фільтру; навести приклади використання.

*Об'єкт дослідження*: інтеграція систем комп'ютерної математики та систем підтримки навчання.

*Предмет дослідження*: процес розробки текстового фільтру системи підтримки навчання Moodle для опрацювання команд системи комп'ютерної математики SageMath.

Використані *методи дослідження*: аналіз державних освітніх стандартів та існуючих аналогів розробки, процес програмної інженерії.

*Результати дослідження*. Розроблений фільтр SageMath надає можливість виконувати код Sage на зовнішньому загальнодоступному сервері SageMathCell, відображати результати виконання на сторінках Moodle без їх перезавантаження за технологією AJAX, є стійким до XSS-атак та готовим для використання у системі Moodle.

*Основні висновки і рекомендації*:

1. Перспективним напрямом розвитку середовища навчання бакалаврів електромеханіки є інтеграція системи підтримки навчання Moodle та системи комп'ютерної математики SageMath.

2. Ефективним засобом убудування моделей системи комп'ютерної математики SageMath у систему Moodle є текстовий фільтр, процес програмної інженерії якого подано у статті.

3. Перспективним напрямом подальших досліджень є використання розробленого фільтру у процесі формування компетентностей бакалаврів

електромеханіки з моделювання технічних об'єктів шляхом вбудовування в навчальні курси, розміщені у системі підтримки навчання Moodle, інтерактивних лабораторних робіт, описаних мовою Sage.

**Ключові слова**: система підтримки навчання Moodle; система комп'ютерної математики SageMath; розробка текстового фільтру; моделювання технічних об'єктів; бакалаври електромеханіки.

### E. O. Modlo[*], S. O. Semerikov[‡]. Development of SageMath filter for Moodle

**Abstract**. *Research goals*: determine the characteristics of the development process, installation, configuration and usage of the filter SageMath for learning support system Moodle.

*Research objectives*: to prove the feasibility of using Moodle system as a tool to support the process of competency formation in technical objects simulation of future bachelors in electromechanical engineering; to analyze existing support tools of technical objects simulation and to identify the ways of it's integration into Moodle; to describe the structure and features of the software implementation of the new SageMath filter for Moodle; to provide the guidance on installing and configuring developed filter; to describe the examples of filter usage.

*Research object*: computer mathematics and learning support systems integration.

*Research subject*: text filter development process for learning support system Moodle to processing the commands of computer mathematics system SageMath.

*Research methods* used: analysis of state educational standards and existing application, software engineering process.

*Research results*. Designed SageMath filter allows to execute the Sage code on the external SageMathCell public server, to view the execution results at the Moodle pages without reloading by using AJAX technology, to stave off XSS attacks and ready for use with Moodle.

*The main conclusions and recommendations*:

1. The perspective direction of learning environment development for bachelors in electromechanical engineering is the integration of learning support system Moodle and computer mathematics system SageMath.

2. An effective tool for embedded a computer mathematics systems SageMath models into Moodle is a text filter. The software engineering process for this filter is presented in the article.

3. Promising area of future research is the use of a developed filter in the process of bachelor's in electromechanical engineering competencies in technical objects simulation by embedding into Moodle learning courses the interactive labs programmed in Sage.

**Affiliation:** Kryvyi Rih Metallurgical Institute, SIHE «Kryvyi Rih National University», 5, Revoliutsiina str., Kryvyi Rih, 50006, Ukraine.

E-mail: modea@mail.ru[*], semerikov@gmail.com[‡].

Розвиток сучасного виробництва неможливий без електроенергетичного забезпечення та використання електромеханічних систем. Фахівці з електроенергетики, електротехніки та електромеханіки забезпечують функціонування тієї основи, на якій надбудовуються усі інші галузі виробництва (включно із електронним).

У зв'язку з цим метою підготовки бакалавра електромеханіки є формування професійно компетентного фахівця, здатного до прогнозування стану та визначення напрямів розвитку електромеханічних систем. Особливої актуальності це набуває сьогодні в умовах інтеграції механічної, електромеханічної, комп'ютерної та системної інженерії на основі концепції STEM (science, technology, engineering, mathematics), що вимагає залучення методів фундаментальних наук у практику роботи інженера-електромеханіка.

Провідним таким методом є моделювання технічних об'єктів, опанування якого забезпечує теоретичне та практичне наповнення фундаментальної, загально та спеціалізовано-професійної підготовки бакалавра електромеханіки. Як було показано у [4], цілеспрямоване формування компетентностей з моделювання технічних об'єктів відбувається у декількох навчальних дисциплінах, провідними з яких є «Загальна фізика», «Теорія автоматичного керування», «Моделювання електромеханічних систем», «Нелінійні та дискретні системи автоматичного керування», «Автоматизація електромеханічних систем» та «Комп'ютерні пристрої в системах автоматизації». Це зумовлює доцільність застосування системи підтримки навчання бакалаврів електромеханіки моделювання технічних об'єктів, спрямованої на формування дослідницької виробничої функції бакалавра електромеханіки, що реалізується через типову задачу діяльності «Проведення дослідних виробничих експериментів (під керівництвом)», що, зокрема, передбачає уміння виконувати опрацювання результатів експерименту; збирати, опрацьовувати і накопичувати вихідні матеріали, дані статистичної звітності, науково-технічні дані тощо; брати участь у дослідженнях систем керування та автоматичного регулювання параметрів.

Однією із найбільш поширених систем підтримки навчання є Moodle. Використання цієї системи разом із відповідними засобами комп'ютерної техніки та спеціальним програмним забезпеченням (насамперед систем комп'ютерної математики) створює умови для формування та розвитку формування компетентностей з моделювання технічних об'єктів.

На сьогодні у світі накопичено значний досвід використання систем підтримки навчання та систем комп'ютерної математики. При автономному використанні систем комп'ютерної математики відсутня автоматизація оцінювання навчальних досягнень, а автономне використання систем підтримки навчання позбавляє можливості виконання дій з моделювання технічних об'єктів безпосередньо у середовищі самої системи. Тому проблема організації взаємодії цих систем є досить актуальною.

Модульна структура системи Moodle надає можливість інтеграції до неї систем комп'ютерної математики шляхом розробки відповідного доповнення до системи (плагіну Moodle), як це показано у [5]. Суттєво спростити цей процес можна шляхом добору засобів, що функціонують у спільному із Moodle хмарному середовищі – таких, як відкрите мобільне математичне середовище SageMath, що надає зручний Web-доступ до великої кількості програмних систем через блокнотний інтерфейс [7]. Автори [1] наголошують, що SageMath дуже зручно використовувати на мобільних Інтернет-пристроях, та наводять приклади мобільних програм, що отримують доступ до серверу SageMath за протоколом XML-RPC.

Одним із інтерфейсів SageMath є SageMathCell [1, с. 921], що надає можливість вбудовування комірок і робочих листів Sage у зовнішні Web-сторінки [7] та відповідає третьому (найвищому) рівню інтеграції системи підтримки навчання та системи комп'ютерної математики [6, с. 189].

Однією із перших спроб такої інтеграції був фільтр Sage для системи Moodle, розроблений авторами [6]. Робота фільтру вимагала:

а) спеціально налаштованого серверу SageMath, встановленого на реальній або віртуальній машині (Amazon Web Sevices, Ulteo Open Virtual Desktop тощо [3]);

б) доступу до серверу SageMath за протоколом XML-RPC;

в) активного фільтру TeX у системі Moodle для відображення результатів запиту до серверу SageMath.

Головним недоліком такого способу інтеграції є необхідність підтримки спеціально налаштованого серверу SageMath адміністратором системи Moodle, що встановлює та налаштовує фільтр. У зв'язку з цим доцільним є розробка нового фільтру SageMath для Moodle, що використовуватиме не приватні, а загальнодоступні сервери, та надаватиме доступ до об'єктів SageMath на стороні клієнта, а не сервера Moodle.

*Мета статті*: опис процесу розробки, встановлення, налаштування та використання нового фільтру SageMath для Moodle.

Для реалізації інтеграції СДН Moodle та СКМ Sage було розроблено фільтр Sage для СДН Moodle. Фільтр надає можливість виконувати код мовою Sage у середовищі Moodle за умови, що він знаходиться у межах тегу `[sage]`…`[/sage]`.

Розглянемо загальну структуру розробленого фільтру. До його складу входять чотири модулі:

1) модуль `version.php` містить відомості про версію плагіну (дата та номер), стабільність версії плагіну (визначена як `MATURITY_STABLE`), його повне ім'я (`filter_sagecell`) та мінімальну версію Moodle, необхідну для роботи плагіну;

2) модуль `settings.php` не містить суттєвого коду – його необхідність визначена вимогами співтовариства розробників Moodle до структури плагінів;

3) модуль `lang/en/filter_sagecell.php` містить ім'я фільтру англійською (`SageCell`);

4) основний модуль `filter.php`.

Даний модуль містить визначення класу `filter_sagecell`, який є похідним від стандартного класу `moodle_text_filter`. У класі перевизначено загальнодоступну функцію `filter`, параметром якої є змінна `text`, що може містити бажану послідовність псевдотегів `[sage]...[/sage]`, усередині якої й буде код мовою Sage.

У системі Moodle одночасно можуть бути активовані декілька фільтрів, що виконуються послідовно при кожній генерації HTML-сторінки. У зв'язку з цим на функцію `filter` накладаються певні швидкісні обмеження. З метою щоякнайшвидшого визначення, чи містить текст, що фільтрується, бажану послідовність псевдотегів, розглядаються три випадки, у яких `text` повертається без змін для опрацювання наступним фільтром:

1) якщо тип даних змінної `text` не є рядковим;

2) якщо `text` є порожнім;

3) якщо до складу `text` не входить рядок `'[sage]'`.

Якщо знайдено рядок `'[sage]'`, виконується виклик функції `preg_replace_callback`, параметрами якої є регулярний пошуковий вираз `'/\[sage](.+?)\[\/sage]/is'`, ім'я функції зворотного виклику для його опрацювання `'filter_sagecell_callback'` та текст, що фільтрується.

Параметром функції зворотного виклику `filter_sagecell_callback` є масив рядків sagecode, другий елемент якого `sagecode[1]` й міститиме код мовою Sage, який користувач увів у вікні текстового редактору Atto, TinyMCE або іншого налаштованого у системі Moodle. При збереженні введеного коду текстовим редактором можуть бути додані ряд HTML-тегів, які повинні бути замінені:

а) теги `<p>`, `</p>`, `<br>`, `<br/>`, `<br />` і т. п. – на символ переведення рядка (`0xA`, або `\n`);

б) нерозривний пробіл (` ` або `\xc2\xa0`) – на звичайний (`\x20`).

Очищений у такий спосіб код все одно може містити XSS-загрози безпеки системи Moodle. Для їх видалення виконується функція `clean_text` ядра системи Moodle. Остаточний код може бути помилковим з точки зору синтаксису Sage, проте безпечним для системи Moodle. Остання дія `filter_sagecell_callback` – його заміна на відповідне звернення до сервера SageMathCell з використанням бібліотек jQuery та embedded_sagecell.js.

Бібліотека embedded_sagecell.js надає можливість створення та убудування у сторінки Moodle комірок SageMath. Для цього створюється обчислювальний текстовий блок (клас блоку `div.compute`), у якому й виконується код Sage (тип мови – `text/x-sage`). Код виконується при завантаженні сторінки Moodle. Для примусового його виконання у SageMathCell користувачу надається кнопка «Evaluate», яка на сторінці Moodle прихована разом із редактором коду Sage.

Повні вихідні коди фільтру розміщено на https://github.com/eugenemodlo/moodle-filter_sagecell. Для встановлення фільтру потрібно завантажити zip-архів із фільтром з GitHub або [2] (рис. 1).

Зауважимо, що архів, завантажений із GitHub, не може бути встановлений у систему Moodle без додаткових налаштувань (рис. 2). Це пов'язано із тим, що ім'я каталогу всередині архіву (moodle-filter_sagecell-master) не відповідає вимогам Moodle до імен каталогів із плагінами (повинно бути sagecell).

Після встановлення фільтр повинен бути активований (переведений до стану «Увімкнуто» за шляхом *Керування сайтом – Модулі – Фільтри – Управління фільтрами*).

Перевірити роботу фільтру можна шляхом уведення команди (послідовності команд) мовою Sage між псевдотегами [sage] та [/sage].

Наведемо у якості прикладу модель трифазної мережі змінного струму (рис. 3).

Фрагмент коду, що містить програмну реалізацію моделі, повинен бути уведений у вікні текстового редактора без використання будь-якого форматування між псевдотегами при створенні нової сторінки курсу, тесту, повідомлення та інших об'єктів, що опрацьовуватимуться фільтром.

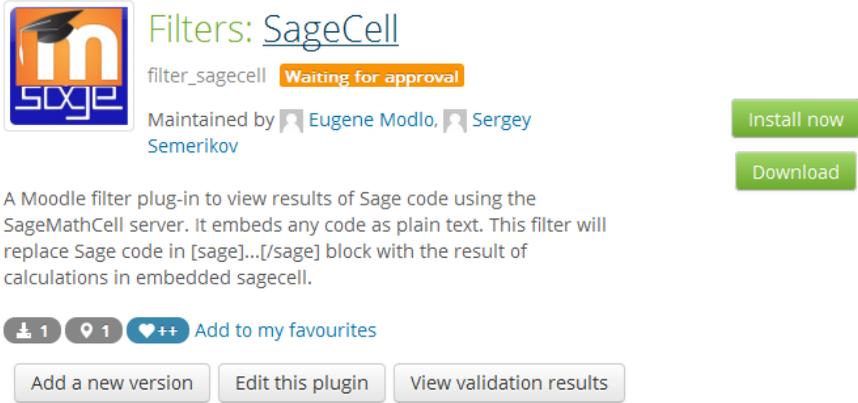

Рис. 1. Сторінка фільтру SageCell у сховищі плагінів Moodle

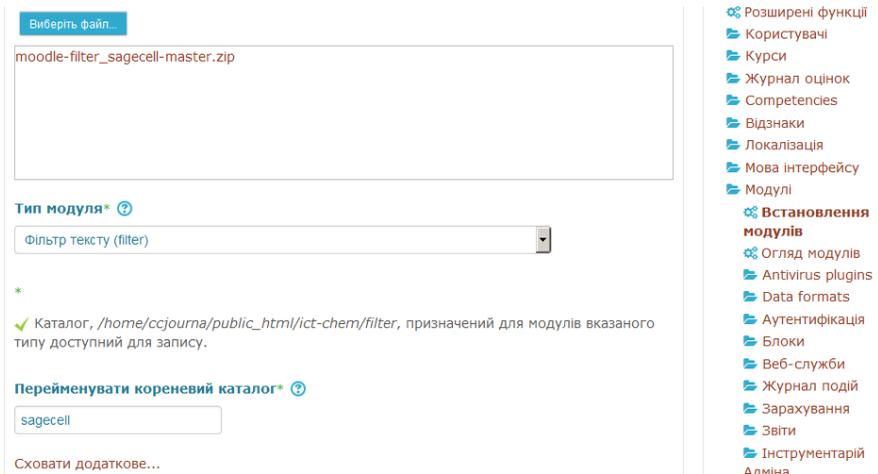

Рис. 2. Додаткові налаштування фільтру SageCell при встановленні із zip-архіву з GitHub

```
[sage]
# модель трифазної мережі змінного струму
var('t')         # часова змінна
A0=380*sqrt(2)   # амплітуда
```

```
w0=2*pi*50      # частота
T0=2*pi/w0      # період відображення
@interact
def model(A=A0,w=w0,T=T0):
    # побудова графіків
    show(plot(A*sin(w*t), t, 0, T, rgbcolor=(1,0,0), \
        thickness=2, legend_label="Phase A") + \
        plot(A*sin(w*t+2*pi/3), t, 0, T, rgbcolor=(0,1,0),\
        linestyle="--", thickness=2, legend_label="Phase B")\
        + plot(A*sin(w*t-2*pi/3), t, 0, T, rgbcolor=(0,0,1),\
        linestyle=":",  thickness=2, legend_label="Phase C"))
[/sage]
```

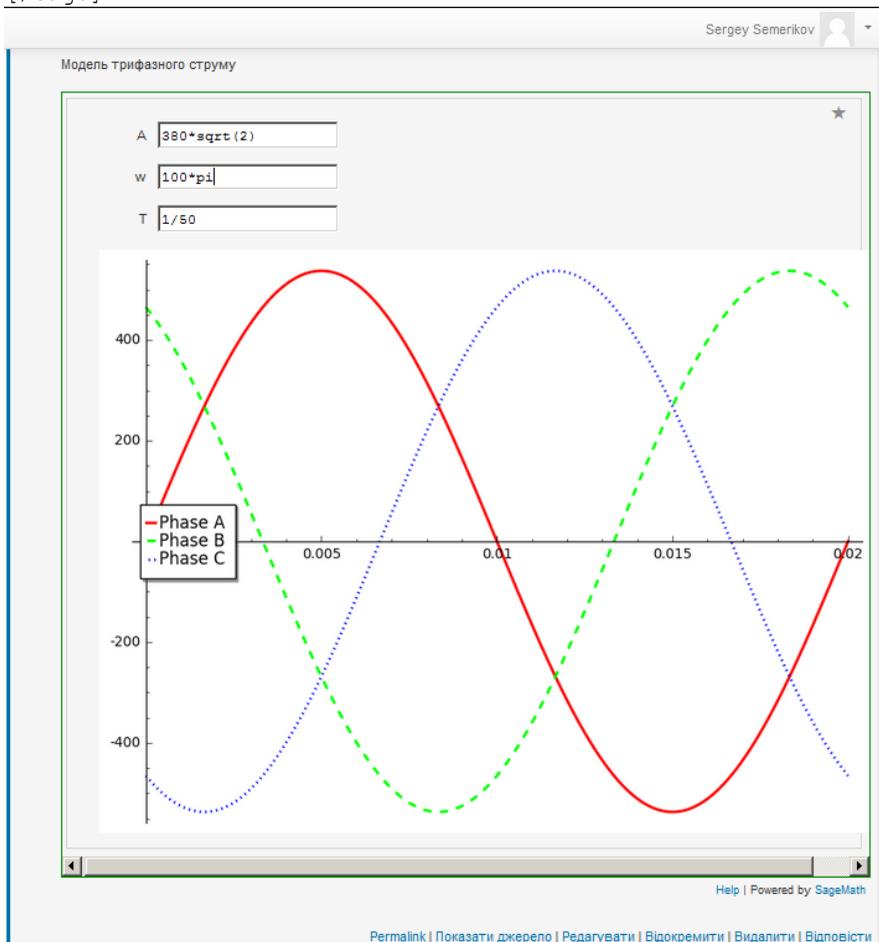

Рис. 3. Модель трифазного змінного струму у системі Moodle

**Висновки:**

1. Перспективним напрямом розвитку середовища навчання бакалаврів електромеханіки є інтеграція системи підтримки навчання Moodle та системи комп'ютерної математики SageMath.

2. Розроблений фільтр SageMath для системи Moodle надає можливість виконувати код Sage на зовнішньому загальнодоступному сервері SageMathCell, відображати результати виконання на сторінках Moodle без їх перезавантаження за технологією AJAX та є стійким до XSS-атак.

3. Перспективним напрямом подальших досліджень є використання розробленого фільтру у процесі формування компетентностей бакалаврів електромеханіки з моделювання технічних об'єктів шляхом вбудовування в навчальні курси, розміщені у системі підтримки навчання Moodle, інтерактивних лабораторних робіт, описаних мовою Sage.

### Список використаних джерел